\documentclass{osa-article}

\journal{oe}
\usepackage[T1]{fontenc}
\usepackage[latin9]{inputenc}
\usepackage{float}
\usepackage{multirow}
\usepackage{amsmath}
\usepackage{amssymb}
\usepackage{graphicx}
\usepackage{wasysym}

\makeatletter

\pdfpageheight\paperheight
\pdfpagewidth\paperwidth

\providecommand{\tabularnewline}{\\}

\usepackage{babel}

\makeatother

\usepackage{babel}
\begin{document}
	%\preprint{APS/123-QED}
	\title{Experimental Demonstration of Underwater Decoy-state Quantum Key Distribution with All-optical Transmission}

	\author{Yonghe Yu,\authormark{1,2} Wendong Li,\authormark{1,2,3} Yu Wei,\authormark{1} Yang Yang,\authormark{1} Shanchuan Dong,\authormark{1} Tian Qian,\authormark{1} Shuo Wang,\authormark{1} Qiming Zhu,\authormark{1} Shangshuai Zheng,\authormark{1} Xinjian Zhang,\authormark{1} and Yongjian Gu\authormark{1,*}}
	
	\address{\authormark{1}Department of Physics, Ocean University of China, Qingdao 266100, China}
	\address{\authormark{2}These authors contributed equally to this work.}
	\email{\authormark{3}liwd@ouc.edu.cn}
	\email{\authormark{*}yjgu@ouc.edu.cn} %% email address is required
	
	%\affiliation{Department of Physics, Ocean University of China.}
	%\date{\today}
	\begin{abstract}
		We demonstrate the underwater quantum key distribution (UWQKD)  over a 10.4-meter Jerlov type III seawater channel by building a complete UWQKD system with all-optical transmission of quantum signals, synchronization signal and classical communication signal. The wavelength division multiplexing and the space-time-wavelength filtering technology are applied to ensure that the optical signals do
		not interfere with each other. The system is controlled by FPGA, and can be easily integrated into watertight cabins to perform field experiment. By using the decoy-state BB84 protocol with
		polarization encoding, 
		we obtain a secure key rate of 1.82Kbps and an 
		error rate of 1.55\% at the attenuation of 13.26dB.  We prove that the system can tolerate the channel loss up to 23.7dB, therefore may be used in the  300-meter-long Jerlov type I clean seawater channel. 
	\end{abstract}

	\section{Introduction}
	
	Quantum key distribution (QKD), a way of generating and distributing secret keys based on quantum physics, is considered unconditionally secure. % from all future advances in mathematics and computing including quantum computers. 
	Following the first BB84
	protocol \cite{BennettBB84}, QKD has been implemented in optical
	fiber \cite{fiber1,fiber2tokyonetwork,fiber3TFqkd} and atmosphere \cite{satalite1,satalite2,satellite3},
	leaving underwater quantum key distribution (UWQKD) a last barrier to conquer.
	In recent years, UWQKD has
	been studied theoretically \cite{UWQKDtheroeyOE,UWQKDtheroeyHINDU,Zhaoshichengbiyelilun} and experimentally \cite{Jin2020,Jinpolar55m,Jintwist55m,JinUWQKD3m,UWQKDtwistcanada,QKDottawariver,Zhaoshichengbiyeshiyan}.
	Most of the experimental studies remain on the level of feasibility research of UWQKD. The polarization of photons was proved experimentally to maintain high fidelity
	through 3 meter water channel and be feasible for UWQKD \cite{JinUWQKD3m}. Twisted photons were also proved feasible for short distance UWQKD \cite{UWQKDtwistcanada}, even in flowing water \cite{QKDottawariver}. In
	subsequent works \cite{Jinpolar55m,Jintwist55m}, quantum tomography was performed, showing that polarized
	photons and twisted photons can survive well through
	a 55 meter underwater channel. Till now, two of the experimental works demonstrated the 
	complete QKD process \cite{Jin2020,Zhaoshichengbiyeshiyan}. BB84 protocol, as the most widely used protocol, was
	completely implemented with polarized photons over a water channel \cite{Zhaoshichengbiyeshiyan}.
	Most recently, decoy-state method was employed in UWQKD \cite{Jin2020}, where the demonstrated distance is up to 30 meters in Jerlov types 2C seawater (equivalent to 345-meter-long clean seawater). 
	
	On the road to the practical application of UWQKD, many 
	works, both scientific and technical, should be performed. One challenge is the requirement of classical optical
	signal in UWQKD. The classical optical signal is widely used to
	send classical information \cite{fiber2tokyonetwork,satalite1,satalite2,satellite3},
	realize time synchronization \cite{fiber2tokyonetwork,satalite1,satalite2,satellite3},
	and finish the task of pointing and tracking in free space QKD \cite{satalite1,satalite2,satellite3}.
	For fiber-based or free space QKD, these  tasks can also be
	finished with the electrical signal or radio wave. 
	However, because of the nature of seawater, electric cable or radio wave cannot be
	used to  finish these tasks underwater, making the wireless optical communication the only choice  for the underwater channel. Moreover, underwater optical communication is even more difficult due to the complexity of the underwater channel and the interactions among the quantum, classical and  synchronization signals.
	Another challenge brought by the underwater channel is that the underwater
	instrument should be designed in small size and with high integration, and therefore should be controlled with elaborately designed field-programmable
	gate array (FPGA) system.
	
	In this work, we successfully implement a complete decoy-state BB84 UWQKD system
	with polarization encoding and all-optical transmission, in which the optical setup, FPGA boards, and software work together. The optical synchronization and optical classical communication are integrated into the system, so the two ends of communication
	do not need any kinds of interactions based on electric cable ors
	radio wave. The lasers and detectors are controlled by one
	FPGA board at each end. 
	%The high level of integration allows this system to be put into a watertight cabin to distribute secret keys in free space underwater environments. 
	We successfully distribute  secret keys
	through 10.4 meter Jerlov type III seawater channel (The channel loss is 13.26dB, equivalent to 170-meter-long Jerlov type I clear seawater channel). The average error rate
	of the sifted keys is 1.55\%, and the average secure key
	rate is 1.82Kbps. The system can tolerate the channel loss 23.7dB, which indicates that
	our system may be used in the  300-meter-long Jerlov type I clean seawater channel.
	
	\section{System Setup and algorithm}
	
	\begin{figure}[b]
		\mbox{%
			\includegraphics[scale=0.043]{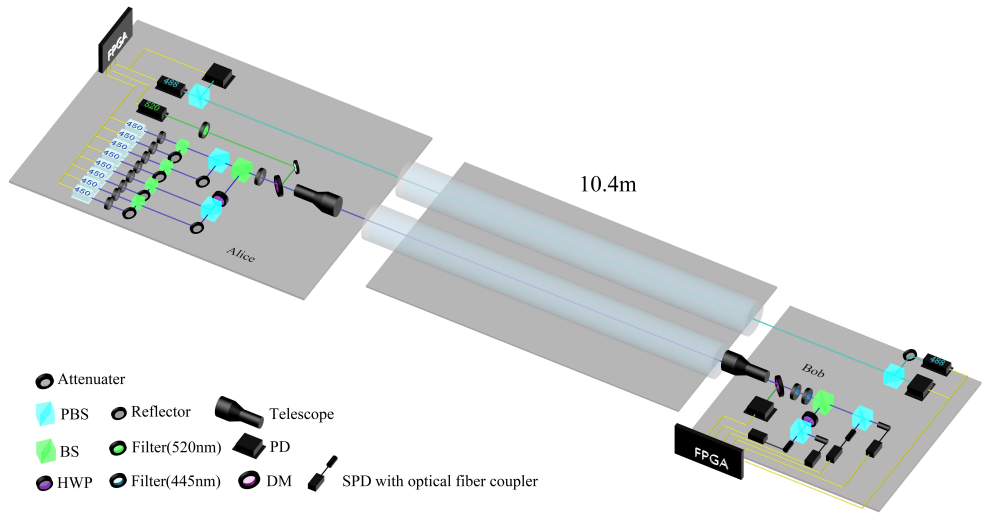}%
		}\caption{\label{fig:epsart-1-1} System setup for the transmitter end (Alice)
			and the receiver end (Bob). HWP: half-wave plate, QWP: quarter-wave
			plate, BS: beam splitter, PBS: polarizing beam splitter, PD: photoelectric
			detector, SPD: single-photon detector made of photomultiplier tube, DM: dichroic mirror. Each FPGA board is connected
			to one personal computer by a USB wire, respectively. Telescopes are used 
			to collimate or collect the quantum and synchronization signals at both ends.}
	\end{figure}

	In our system shown in Fig. \ref{fig:epsart-1-1}, at the Alice end, we use eight blue lasers ($450$nm) to implement the three-intensity decoy-state polarization-encoding BB84 protocol because of the lack of effective polarization modulators in the blue-green region. We use one green laser ($520$nm) to generate the synchronization signal, another laser ($488$nm) to transmit classical communication signal, and one photoelectric detector~(PD) to detect the classical communication signal. 
	
	At the Bob end, the quantum signals	are detected with four single-photon detectors (SPD) made of photomultiplier tubes. One laser ($488$nm) is used to transmit classical communication signal, and two PDs are used to detect the synchronization signal and the classical communication	signal.
	
	The dichroic mirrors are employed to combine the quantum and the synchronization signals at the Alice end and to split them at the Bob end.
	Two water pipes are used as the underwater channels. One is used for the propagation of quantum signals and synchronization signal, and the other is for the classical communication signal. 
	All lasers, PDs, and SPDs are controlled
	by FPGA circuit boards at both ends. The FPGA circuit boards also perform key-sifting and transmit the sifted keys to the personal computers (PCs).
	The software in the PCs completes error-correction, error
	checking, and privacy amplification. With this UWQKD system,
	real-time secret keys can be generated.

	\subsection{Quantum signal source and polarization encoding}

\begin{figure}
		\centering
		\mbox{%
			\includegraphics[scale=0.14]{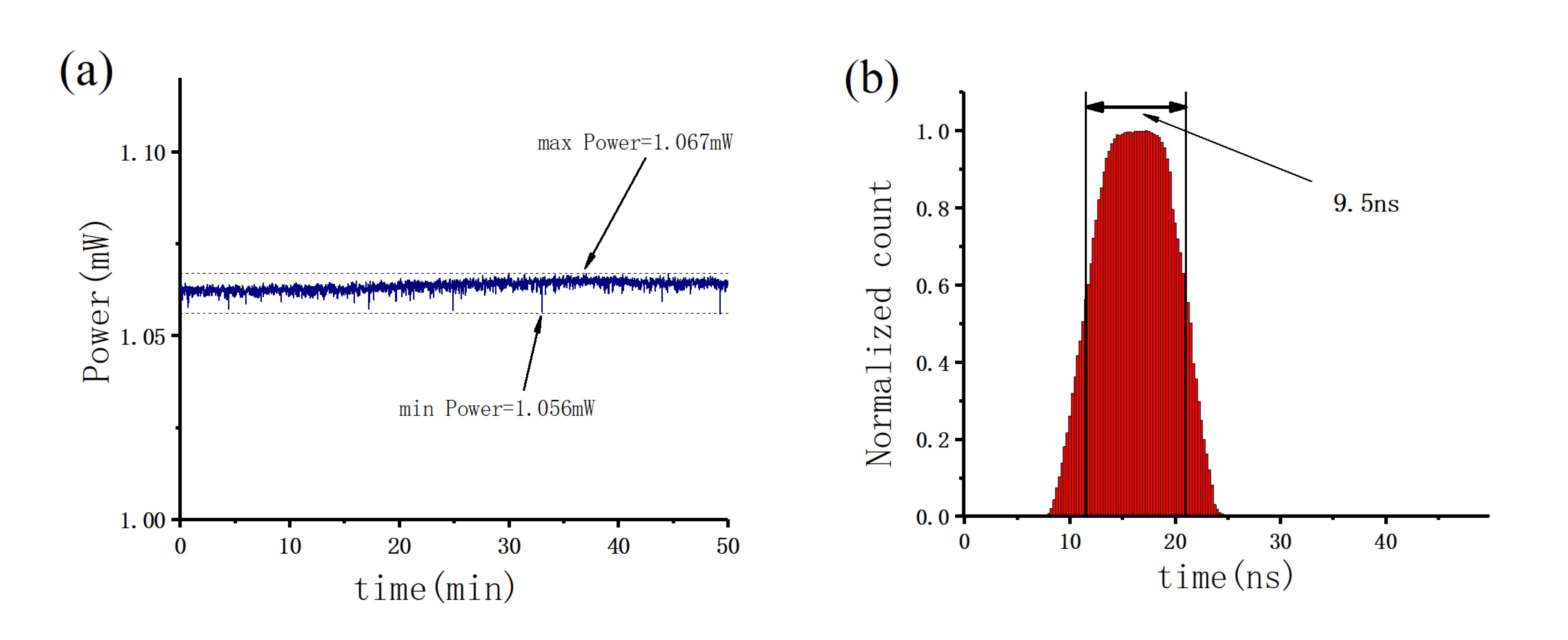}%
		}\caption{\label{fig:epsart2-1} Performance tests of the quantum lasers. (a) The power stability test in 50 minutes. The power variation is less than 0.011mW(about 1\%). (b) The temporal shape of the quantum signal pulse. SPD is used 
			to count the photons to obtain the shape. The eight quantum lasers
			have almost completely the same temporal shapes.}
	\end{figure}

	The central wavelength of quantum signal lasers is $449.5$ nm with the deviation less than $0.5$nm, and the power
	stability is less than $1$\% within 50 minutes as shown in Fig. \ref{fig:epsart2-1}(a). These eight lasers are organized into four groups, each group corresponding to one of the four polarizations (horizontal (H),
	vertical (V), $45{^\circ}$ (P), $135{^\circ}$ (M)), and the two lasers in one group generating the pulses of signal state and the decoy-state through different attenuations required by the decoy-state method. The beam splitter~(BS) 
	and reflector are used to combine the signal state and
	the decoy state in a group. Then the PBSs, HWP and BS are used to encode the pulses into the four polarizations and combine them together.
	
	The control format for the quantum signal lasers is shown in Table~\ref{tab:table1-1}. We use four random number generator~(RNG) chips to generate the random
	numbers (denoted by bit3, bit2, bit1 and bit0), which are fed into the FPGA to
	control the operation of the eight quantum signal lasers. The FPGA modulates the lasers with the repetition frequency $20$MHz, and the measured pulse	width is $9.5$ns as shown in Fig. \ref{fig:epsart2-1}(b). 
	
	\begin{table}[b]
		\caption{\label{tab:table1-1} The control format for the quantum signal lasers}
		\centering
		
		\begin{tabular}{lc}\hline\hline
			Bit3 and bit2  & Polarization\tabularnewline\hline
			00  & H\tabularnewline\hline
			01  & V\tabularnewline\hline
			10  & P\tabularnewline\hline
			11  & M\tabularnewline\hline\hline
		\end{tabular}
		\begin{tabular}{|lc}\hline\hline
			%Random Number(bit1 and bit0)  & Pulse Intensity\footnote{The mean photon number of each state can be adjust by the attenators}\tabularnewline
			Bit1 and bit0  & State\tabularnewline\hline
			00  & vacuum state\tabularnewline\hline
			01  & weak decoy state\tabularnewline\hline
			10  & signal state\tabularnewline\hline
			11  & signal state\tabularnewline\hline\hline
		\end{tabular}%
		
	\end{table}

	\subsection{Synchronization, wavelength division multiplexing~(WDM) and filtering}
	
	In our system, the repetition frequency of the synchronization signal
	is $5$MHz and the pulse width is $30$ns. We use the wavelength division multiplexing~(WDM) technology to combine and split the synchronization signal and the quantum signals. Specifically,
	a dichroic mirror with edge at 480nm is used at the Alice end to combine the signals by letting the quantum signals (450nm) pass through and reflecting the synchronization signal (520nm), and another dichroic mirror is used to split them at the Bob end. 
	%The dichroic mirrors can let light with wavelength	smaller than $470$nm pass through and reflect light with wavelength larger than $490$nm. 
	Our test shows that the ratio of the 520nm laser passing through the dichroic mirror is less than 0.1\%.
	
	At the Bob end, the PD converts the optical synchronization signal into the electrical signal. In order to  match the quantum signal, the electrical signal is delayed by a high-precision
	programmable delay chip, whose delay precision is $250$ps with a maximal delay $64$ns, and then sent to the  FPGA.	The FPGA converts the signal into the gate signal of repetition frequency $20$MHz, and the gate width is set to $10$ns. We design a self-adapting algorithm to precisely match the quantum signal. 
	
	To reduce the influence of the synchronization signal and background noise, 
	we perform filtering in space, time and wavelength domains. In time domain, we use the FPGA program to omit clicks of the SPDs out of the time gates. The clicks coming from the synchronization signal will also be omitted by turning off the  gate when the synchronization pulses arrive. In wavelength domain, we use two filters behind the dichroic mirror at the Bob end. The spectral width of the filters is $20$nm, and the central wavelength is $445$nm. Our test shows that these two filters perform well in suppressing the background noise. 
	In space domain, the field of view (FOV) is limited mainly by the fiber coupler and the aperture is limited by the telescope. The FOV of the fiber coupler is about 0.14 degrees and the diameter of the telescope is 5cm. When we use SPDs directly  without the fiber coupler and fiber, the noise caused by the classical signal and the background light makes the error rate extremely high, and the UWQKD experiment cannot be performed successfully.
	
	A big challenge in our early experiment is that the count rate of the SPDs was always high when the optical synchronization signal was open, even when many 445nm filters were used. We speculate that the synchronization laser can also send photons in $450$nm. To protect the SPDs, we use a $520$nm filter
	at the Alice end just behind the synchronization laser and solve the problem perfectly. 
	
	\subsection{Optical classical communication}
	To build a system with all-optical transmission, the classical communication in the UWQKD system is realized by optical signal with the open-off-keying~(OOK) modulation~\cite{ook1,ook2}. The wavelength of the classical communication signal is $488$nm, and the baud rate is $20$MHz. Horizontal and vertical polarizations are employed at two ends, respectively, so as to use the same channel to realize the duplex optical classical communication via the two PBSs.
	
	\subsection{FPGA control}
	
	\begin{figure}
		\centering
		\mbox{%
			\includegraphics[scale=0.34]{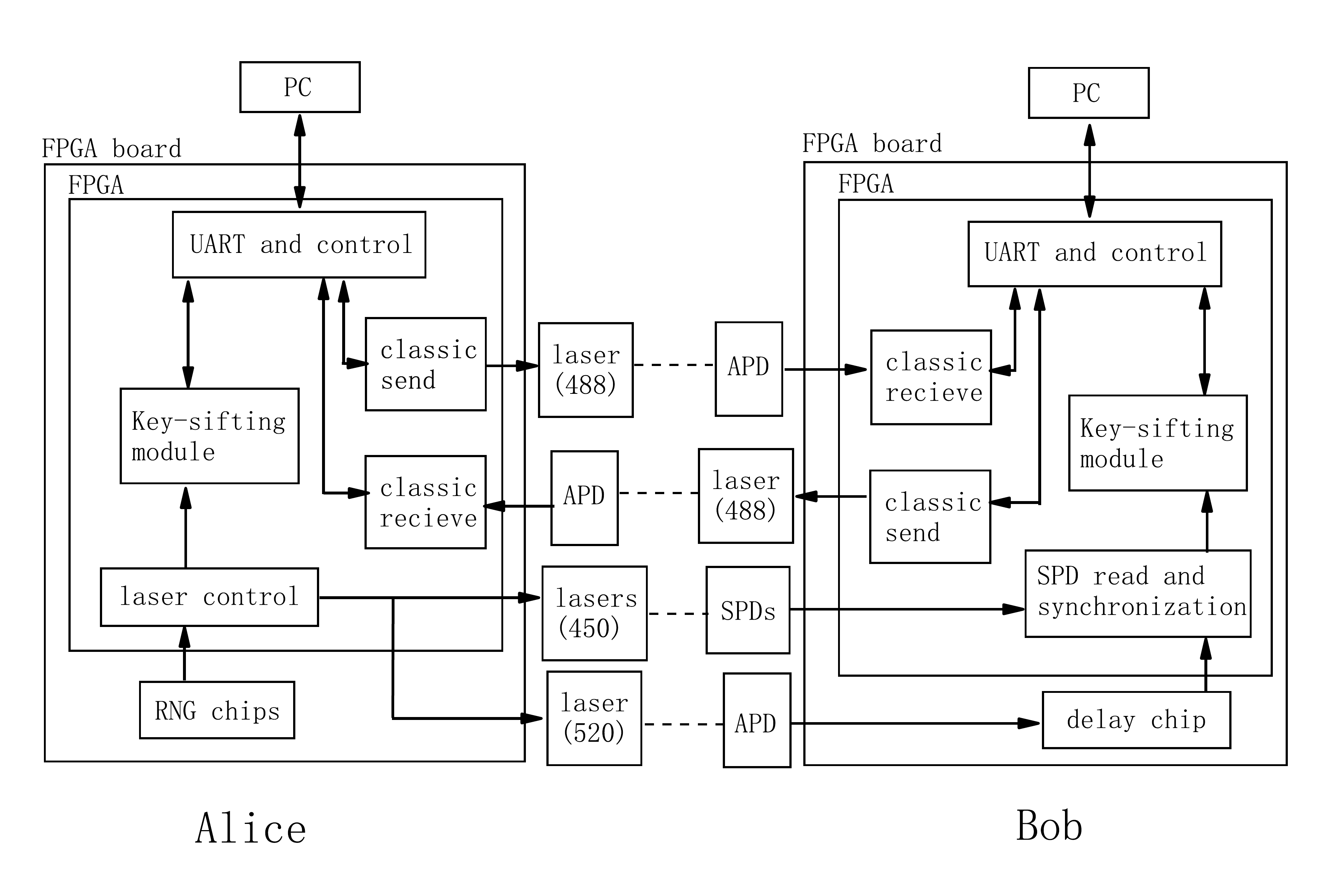}%
		}\caption{\label{fig:epsart7} FPGA diagrams at the Alice and Bob ends. The dotted lines represent the optical links through seawater.}
	\end{figure}
	
	FPGA is responsible for the control of the decoy-state UWQKD system, including polarization encoding and detection, key-sifting, classical information between Alice and Bob and communication with PCs. As shown in Fig. \ref{fig:epsart7}, at the Alice end, the laser control module reads random numbers from the RNG chips and converts them into quantum and synchronization laser control pulses to generates keys, and then sends the keys to the key-sifting module. The key-sifting module temporarily stores the keys and completes the key-sifting with the aid of classical communication with Bob. 
	The UART and control module sends the sifted keys to the PC. Both the key-sifting module and the PC can send information to "classic send" module and receive information from the "classic receive" module through the UART and control module. At the Bob end, the "SPD read and synchronization" module controls the delay chip to match the synchronization and the quantum signals, completes the single-photon detections to generates the raw keys, and sends the keys to Bob's key-sifting module. Other modules are similar to the modules at the Alice end.

	\subsection{Post-processing}
	
	In QKD process, post-processing is required to generate the secure keys, including key-sifting, error correction, error checking, and privacy amplification \cite{Charles1988Privacy,Privacy2}. Firstly, in the key-sifting, Alice's and Bob's FPGAs exchange information to sift and preserve the keys which have correct bases. In this process, Alice announces 20\% of the keys through the classical channel to Bob for estimating the error rate. At the Bob end, the key rates and error rates of signal state, decoy state and vacuum state are sent to PC. 
	Secondly, the rest of the sifted keys participate in the error correction, which is realized by low-density parity check code (LDPC). In our system, the code length of LDPC code is 9216 and the coding efficiency is 3/4B, following the criteria of IEEE802.16e. 
	Thirdly, error checking \cite{zhangchunmei} and privacy amplification are performed to make sure the final keys are the same and safe. The total number of the secure keys is calculated and the sifted keys are compressed into the secure keys.
	The Toeplitz \cite{toeplitz} matrix is used to complete the hash function in error checking and privacy amplification. To reduce the impact of the finite-size effect, we conduct the privacy amplification for every 256 groups of LDPC decoding data, which means that roughly 1.7Mbits of keys participate in privacy amplification once. Due to the use of Fast Fourier transform (FFT), our system takes only 2 seconds to finish each error checking and privacy amplification.

	\section{Experiments and Results}
	
	\subsection{Characterization of the underwater channel and the optical system}
	
	In the experiment, the seawater we used is collected by Dongfanghong Research Ship of Ocean University of China. 
	The length of the water channel is 10.4m and  the attenuation coefficient is $0.293m^{-1}$ (Jerlov type III seawater), thus the overall attenuation
	of the $10.4$m channel is $13.26$dB (450nm). The salinity of the water is $32\permil$ and the temperature is $13^{\circ}C$. The total photon collecting efficiency of Bob is 54.9\%, and
	the quantum efficiency
	of the SPD is 20\%. Therefore the overall attenuation of Bob is
	$9.59$dB.

	%\subsection{Quantum state tomography experiment in 10.4m underwater channel}
	\subsection{Quantum tomography}
	
	We perform quantum tomography over  the 10.4m underwater channel at the single-photon level before the QKD experiment. In the experiment, Alice's FPGA controls the quantum lasers to send four polarization states, and Bob's FPGA counts the photons detected by the four single-photon detectors to calculate the outcoming density matrices. The ideal density matrices of the four polarization states in this work are: 
	
	\begin{equation}
	\begin{aligned}
	\rho^{H}_{ide}&= \text{\ensuremath{\left[\begin{array}{cc}
			1 & 0\\
			0 & 0\end{array}\right]}},~&
	\rho^{P}_{ide}&= \text{\ensuremath{\left[\begin{array}{cc}
			1/2 & 1/2\\
			1/2 & 1/2\end{array}\right]}},\\
	\rho^{V}_{ide}&= \text{\ensuremath{\left[\begin{array}{cc}
			0 & 0\\
			0 & 1\end{array}\right]}},~&
	\rho^{M}_{ide}&= \text{\ensuremath{\left[\begin{array}{cc}
			1/2 & -1/2\\
			-1/2 & 1/2\end{array}\right]}}. 
	\end{aligned}
	\end{equation}

\begin{figure}[b]
		\mbox{%
			\includegraphics[scale=0.13]{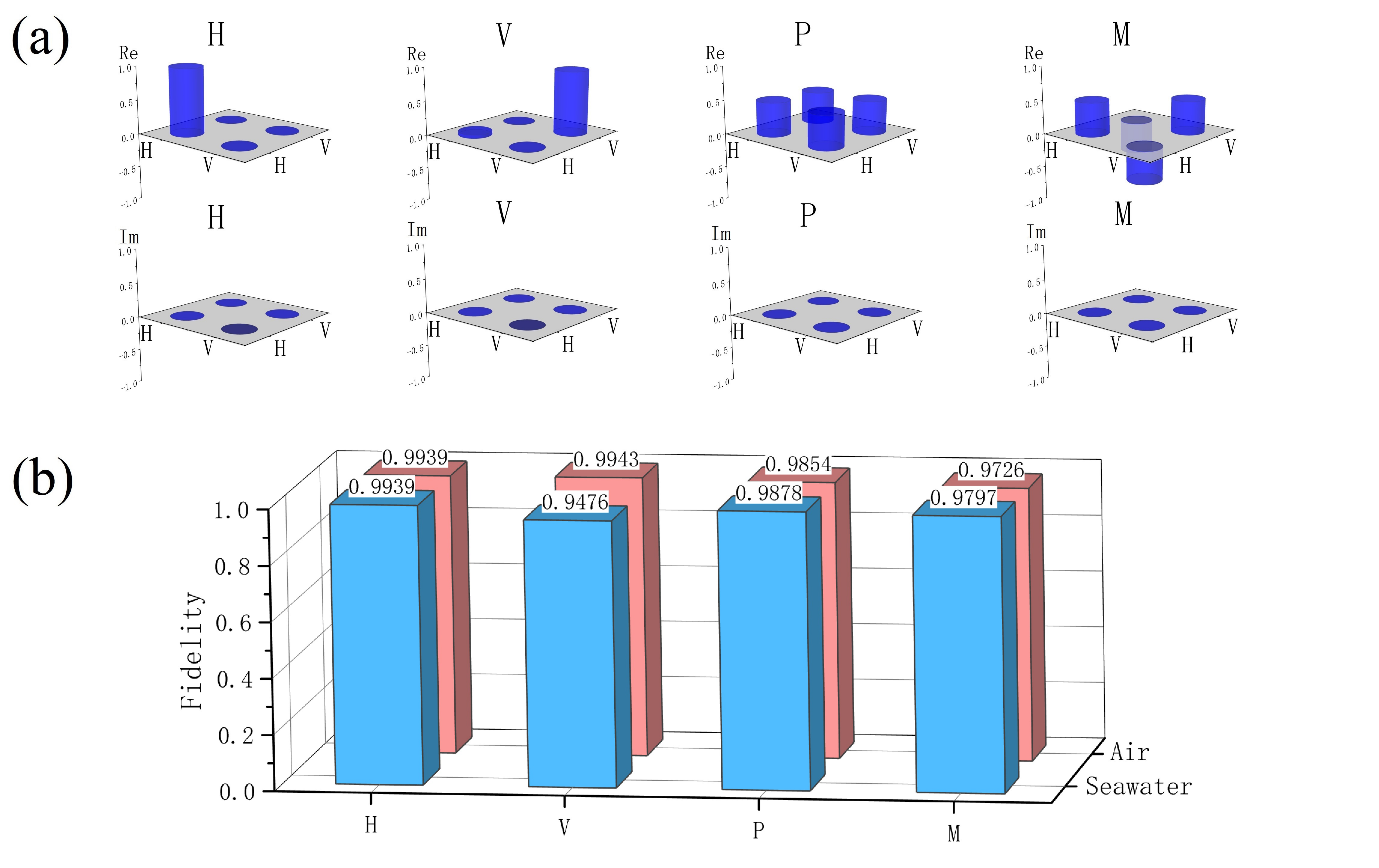}%
		}%
		\caption{\label{fig:epsart-1} The results of quantum tomography of the polarization states. (a) The measured density matrices of the four polarization states, with Re and Im being their real parts and imaginary parts, respectively. (b) The fidelities of the four polarization states.}
	\end{figure}

	The results of the tomography are shown in Fig. \ref{fig:epsart-1}(a). 
	From the density matrices, the state fidelity can be calculated according to $F_{s}=\left[tr\left(\sqrt{\sqrt{\rho_{mea}}\rho_{ide}\sqrt{\rho_{mea}}}\right)\right]^{2}$, where $\rho_{mea}$ are the measured density matrices of the four polarization states. In Fig. \ref{fig:epsart-1}(b), the fidelities of the four polarization states of photons passing through the seawater channel and air are shown for comparison. It can be obtained that the average fidelities are $0.9773$ for the water channel and $0.9866$
	for the air. In addition, the count rates of the background noise and dark count of the SPDs are, respectively, below 200Hz and 50Hz, far below the single photon count rate (over 200KHz) of the signal. These results prove that seawater has slight effect on the polarization, which indicates that polarization-encoding is fit for UWQKD.

	%\subsection{Decoy State BB84 protocol QKD in underwater channel}
	\subsection{Decoy-state UWQKD}
	
\begin{figure}
		\centering
		\mbox{%
			\includegraphics[scale=0.23]{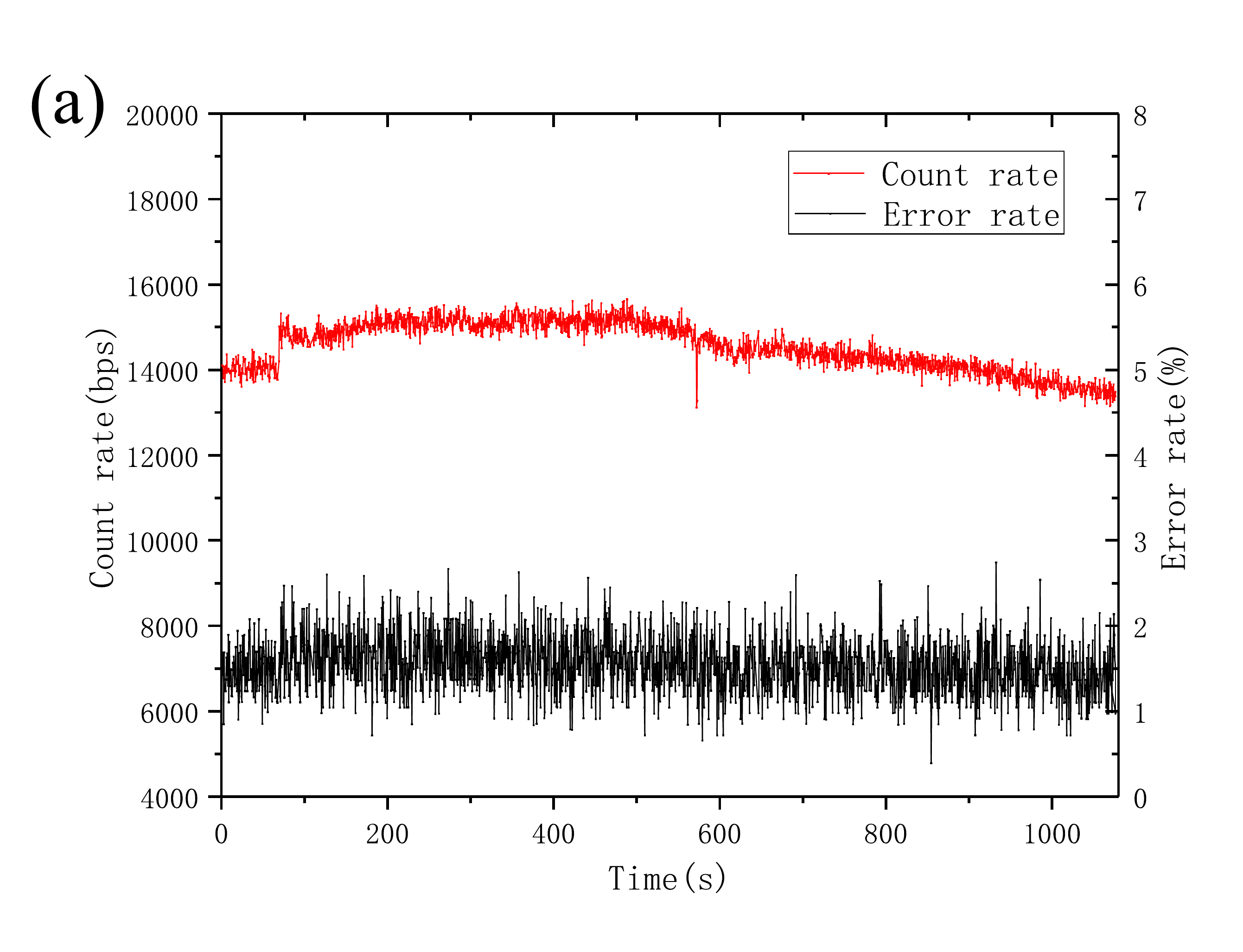}\includegraphics[scale=0.23]{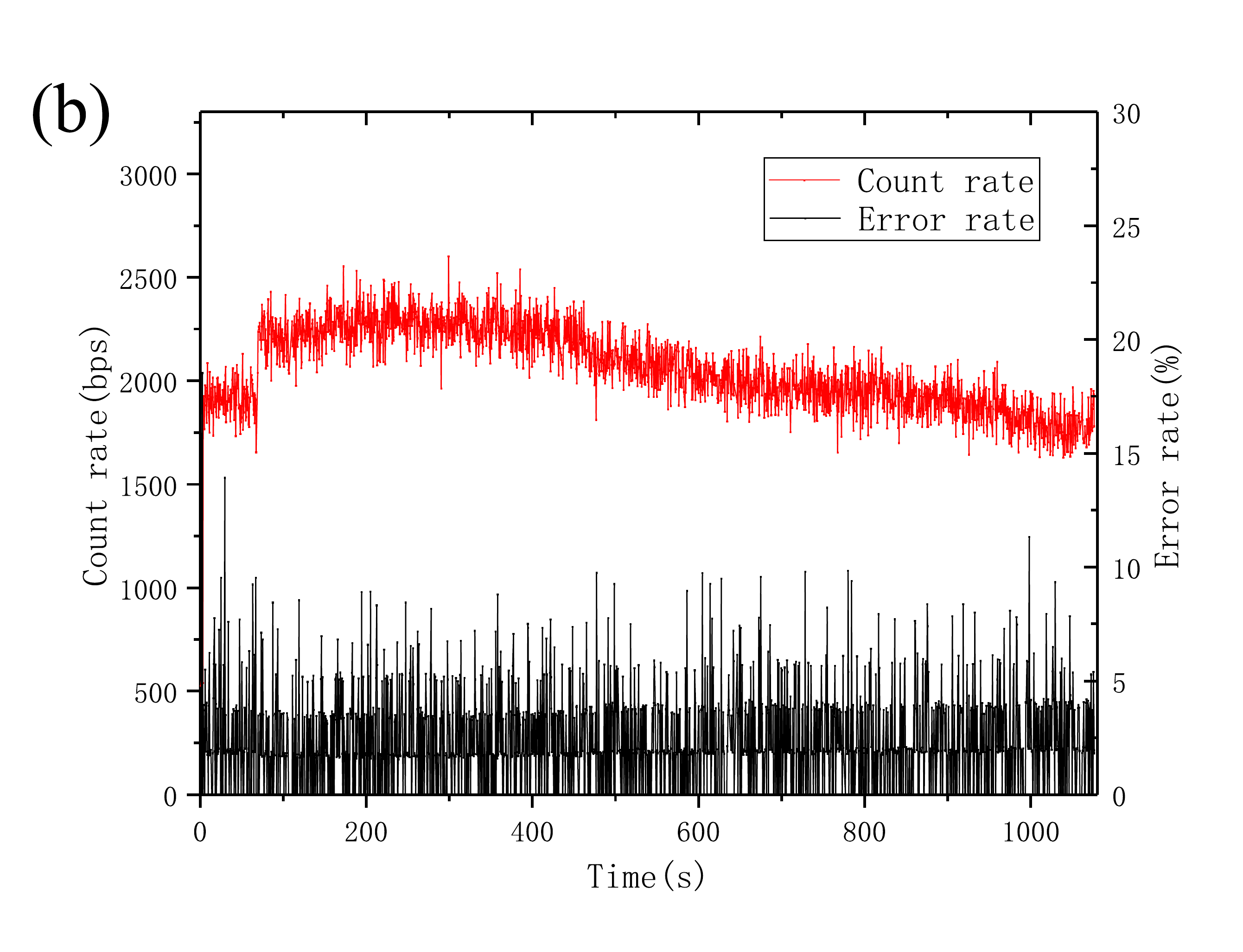}
		}
		\caption{\label{fig:epsart3} The experimental results of the UWQKD experiment through the 10.4m seawater channel. 
		(a) The real time count rates and error rates for the signal state. 
		(b) The real time count rates and error rates for the decoy state.}
	\end{figure}

	We demonstrate complete key distribution with our decoy-state BB84 QKD system through the 10.4m seawater channel. In the QKD experiment, the average photon numbers per pulse of the signal state and the decoy state are 0.8 and 0.1 respectively. After the PCs at the both ends send the start order to their FPGAs, the FPGAs execute the decoy-state protocol and report the sifted keys, the sifted key rates and error rates to PCs. PCs calculate the secure key rate and generate the secure keys. The results of a $1077$ second experiment are shown in Fig.~\ref{fig:epsart3}.
	
	Fig.~\ref{fig:epsart3}(a) and (b) show the (sifted) count rates and error rates of the signal state and decoy state. It can be seen that the average (sifted) count rate of the signal state is more than $14$Kbps and the average error rate of the signal state is 1.55\%. This number is low enough for the LDPC code to carry out the error correction successfully. In fact, Alice and Bob can obtain completely the same final keys. We can also see that the count rates of the decoy state are about $2$Kbps. The average error rate of decoy state is only $2.35$\%  despite the large fluctuation caused by its smaller count rates, because there are a considerable number of zero points in the error rates of decoy states.

	The real-time secure key rate is not shown in Fig.~\ref{fig:epsart3}, for the reason that
	the privacy amplification is conducted for every $1.7$Mbits sifted
	keys, which means that the system needs $2$ minutes to accumulate
	the sifted keys for one privacy amplification. Instead, we show
	the (sifted) count rates and error rates in Fig.~\ref{fig:epsart3} and in Table \ref{tab:table1-1-1}.

	\begin{table*}
		\caption{\label{tab:table1-1-1} Average count rates and error rates.}
		\centering
		\begin{tabular}{ccccc}\hline\hline
			\multirow{2}{*}{Average count rates (bps)} & $Q_{u}$ & $Q_{v}$ & $Q_{0}$ & $Q_{1}$\tabularnewline\cline{2-5} 
			& 14564.7 & 2059.1 & 16.7 & 7205.9\tabularnewline\hline
			\multirow{2}{*}{Average error rates} & $E_{u}$ & $E_{v}$ & $E_{0}$ & $e_{1}$\tabularnewline\cline{2-5} 
			& 1.55\% & 2.35\% & 50.44\% & 2.25\%\tabularnewline\hline
			\multirow{2}{*}{Average secure key rate (bps)} & \multicolumn{4}{c}{$R_{SKR}$}\tabularnewline\cline{2-5} 
			& \multicolumn{4}{c}{1823.4}\tabularnewline\hline\hline
		\end{tabular}
	\end{table*}

	As shown in Table \ref{tab:table1-1-1}, thanks to the adoption of the time gates, the average count rate of vacuum state $Q_{0}$ is $16.7$bps, much lower than the
	background noise in pre-experiment without time gates, which means our temperal filtering works well. The count rate $Q_{1}$ and error rate $e_{1}$ of single photons can be estimated by the count rate $Q_{u}$ and the error rate $E_{u}$ of signal state, the count rate $Q_{v}$ and the error rate $E_{v}$ of decoy state, the count rate $Q_{0}$ and the error rate $E_{0}$ of vacuum state according to the decoy state method.
	The results are that the count rate of single photons is 7205.9bps and the error rate of single photons is 2.25\%.

	In the experiment, $1963762$bits secure keys are generated in 1076 seconds, and the secure key rate $R_{SKR}$ is 1823.4bps. The system do not need to stop single-photon distribution to do the privacy amplification. The privacy amplification is conducted just when the
	sifted keys reach $1.7$Mbits. Secure keys are generated in situ  and in real-time.

	To transmit quantum and synchronization signals in one channel and classical signals in another channel, we use two water pipes filled with the same seawater. However, as mentioned above, even if two water pipes are used, the influence of classical signal on the quantum signal cannot be ignored unless the space-time-wavelength filtering are applied.
	
	There are still some issues worthy to be considered for the practical application of UWQKD. For example, the acquisition pointing tracking (APT) in the underwater turbulence and strong attenuation environment. The addition of the APT system requires the introduction of additional classical optical signals, and our experiments have proved the feasibility of adding classical signals in the quantum key distribution system.

	%\section{Discussion of the system performance in higher loss condition\label{sec4}}
	\section{Numerical Simulation of the System Performance\label{sec4}}
	
	As shown above, our system works well under the condition of
	$10.4$m seawater and $13.26$dB attenuation. In fact, our
	system can tolerate much higher level of attenuation. Next we show the simulation results of the performance of our system under
	different attenuations. The simulation parameters are shown in Table 
	\ref{tab:table3}.
	
	\begin{table}[ht]
		\caption{\label{tab:table3} System parameters of numerical simulation}
		
		\centering
		\begin{tabular}{cccccc}\hline\hline
			$F$(MHz)  & $u/v$  & $\eta_{opt}$(dB)  & Ratio of the three states  & $Q_{0}$(bps)  & $e_{det}$\tabularnewline\hline
			20  & 0.8/0.1  & 9.59  & 2:1:1  & 16.7  & 1.5\%\tabularnewline\hline\hline
		\end{tabular}

	\end{table}
	
\begin{figure}[ht]
		\centering
		\mbox{%
			\includegraphics[scale=0.32]{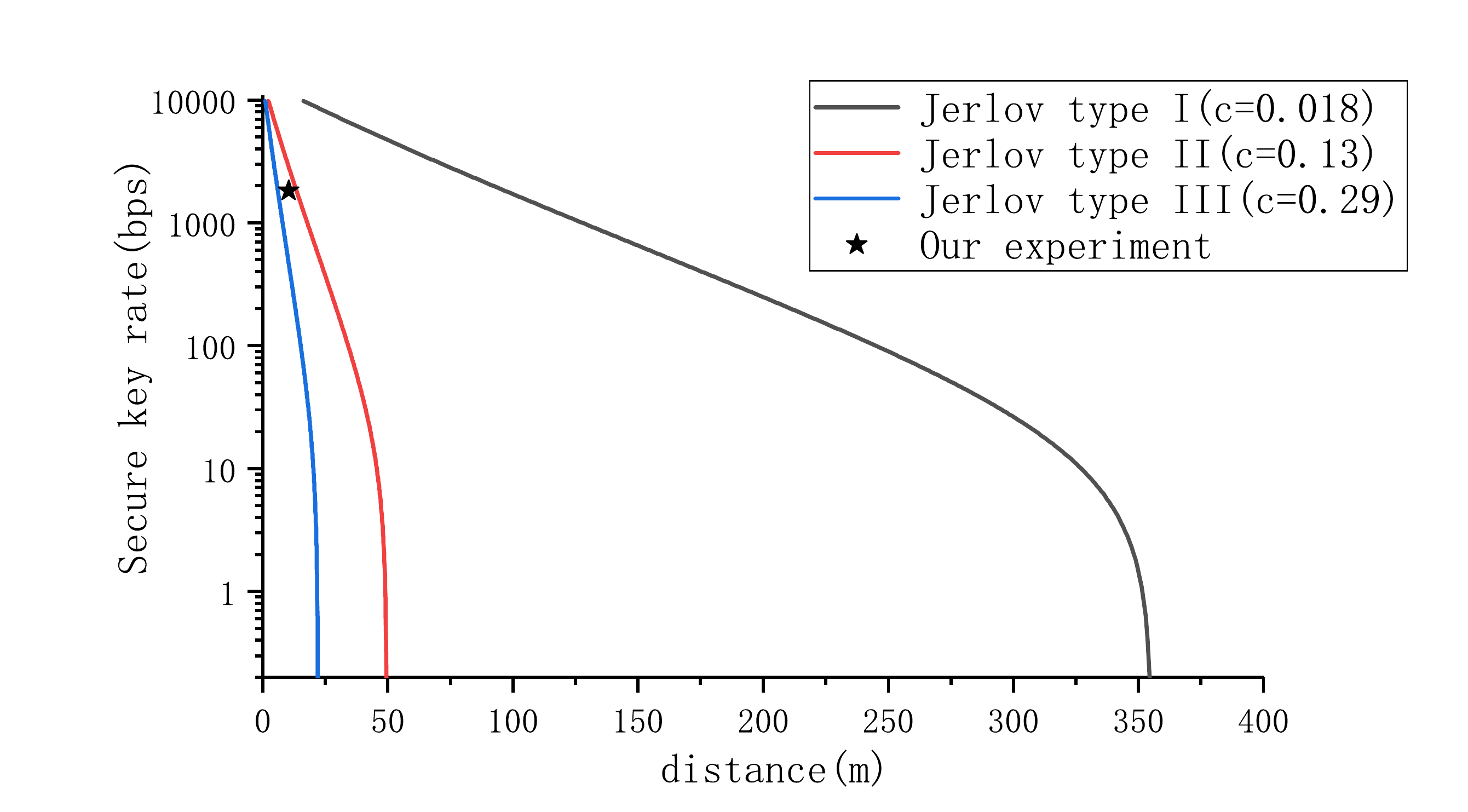}%
		}\caption{\label{fig:epsart6} Secure key rates of our system in different type of water as a function of distance. c is the attenuation coefficients of the seawater.}
	\end{figure}

	The simulation parameters are taken from our experiments.
	Specifically, $F$ is the repetition frequency of the quantum signals, $u$ and v
	are the mean photon numbers per pulse for the signal state and the decoy state,
	$\eta_{opt}$ is the	photon loss caused by the optical setup and the single-photon detectors at the Bob end, $e_{det}$ is the error rate caused by the degradation of the polarization, including the influence of the channel and the optical 
	elements.
	For simulation purpose, $e_{det}$ is set to $1.5$\% without loss of generality.
	For our system, the secure key rate $R_{SKR}$ can be estimated by the following formula~\cite{decoyPRA,UWQKDtheroeyOE,Jin2020}:
	\begin{equation}
	R_{SKR}\geqslant Q_{1}[1-H_{2}(e_{1})]-Q_{u}R,\label{eq:1}
	\end{equation}
	where $H_{2}(x)$ is 
	the binary Shannon information function, $R$ is the unsafe fraction in sifted keys caused  by LDPC. The count rate of single photons $Q_{1}$, the error rate of single photons $e_{1}$, and the count rate of signal state $Q_{u}$ are calculated according to the decoy-state method \cite{decoyPRA}. 
	%In our system, the number of secure keys is equal to the difference between the number of single photons which are not eavesdropped and the information revealed by error correction. The information revealed by error checking and privacy amplification is not included into Eq. (\ref{eq:1}).
	The check code produced by the LDPC encoding will be sent to Bob by the classical channel as the side information. In the worst condition, we assume all of the information sent through the classical channel is eavesdropped and R equals to $1/3$ in our system.
	For different type of water,
	the results of the secure key rates as a function of the distances are shown
	in Fig.~\ref{fig:epsart6}.

	The random numbers for generating the Toeplitz matrix are produced by the
	same seeds in the PCs of Alice and Bob. This process does not require
	classical message interaction, while the information revealed by error
	checking is very few~(8bits for one error checking and privacy amplification in our system). Thus the information revealed by error checking and privacy amplification is not included into Eq. (\ref{eq:1}).

	Fig. \ref{fig:epsart6} shows that our system can still produce secure keys in $300$m Jerlov type I seawater, where the attenuation is 23.7dB. In this case, the sifted-key rate $Q_{u}$ is 1200.1bps and the error rate of signal state $E_{u}$ is 2.15\%, so the secure key rate $R_{SKR}$ is 27.4bps. %fe=1.16:Rskr=219.2
	In real post-processing and in simulation here, we adopt the  LDPC error correction scheme instead of using the ideal error correction efficiency (1.16) directly. If we use the latter in Eq. (\ref{eq:1}), the $R_{SKR}$ can reach 219.2bps. The LDPC error correction scheme we adopted consumes more secure keys than the ideal case, 
	but it represents the real situation. In the case of $300$m Jerlov type I seawater, taking the attenuation of Bob's optical setup $\eta_{opt}$ (9.59dB) into account, our system can tolerate the overall attenuation $33.3$dB.

	\section{Conclusion}
	
	We develop an underwater decoy-state quantum key distribution system, and perform the quantum key distribution experiment through 10.4m Jerlov type III seawater channel. The average error rate
	of the sifted keys is 1.55\%, and the average secure key
	rate is 1.82Kbps. Instead of using radio waves or electrical cable which are not fit for the underwater channel, the two ends of our system completely use optical signals to communicate, which makes the system able to work in the underwater environment. 
	The system is controlled by FPGA, and can be easily integrated into watertight cabins to perform the field experiment. 
	Furthermore, we prove that our system is able to tolerate the channel attenuation up to 23.7dB, and therefore can be used in the  300-meter-long Jerlov type I clean seawater channel.

	\begin{backmatter}
		\bmsection{Funding}
		This work was supported by the National Natural Science Foundation
		of China (Grants No. 615705180, No. 61701464, No. 12005212, and No. 201861012).
		
		\bmsection{Acknowledgments}
		The authors thank Long-wen Zhou, Zhi-min Wang for their help in preparing the paper.

	\end{backmatter}

	\appendix
	
	\section{Appendixes}
	\begin{figure}[ht]
		\centering
		\mbox{%
			\includegraphics[scale=0.04]{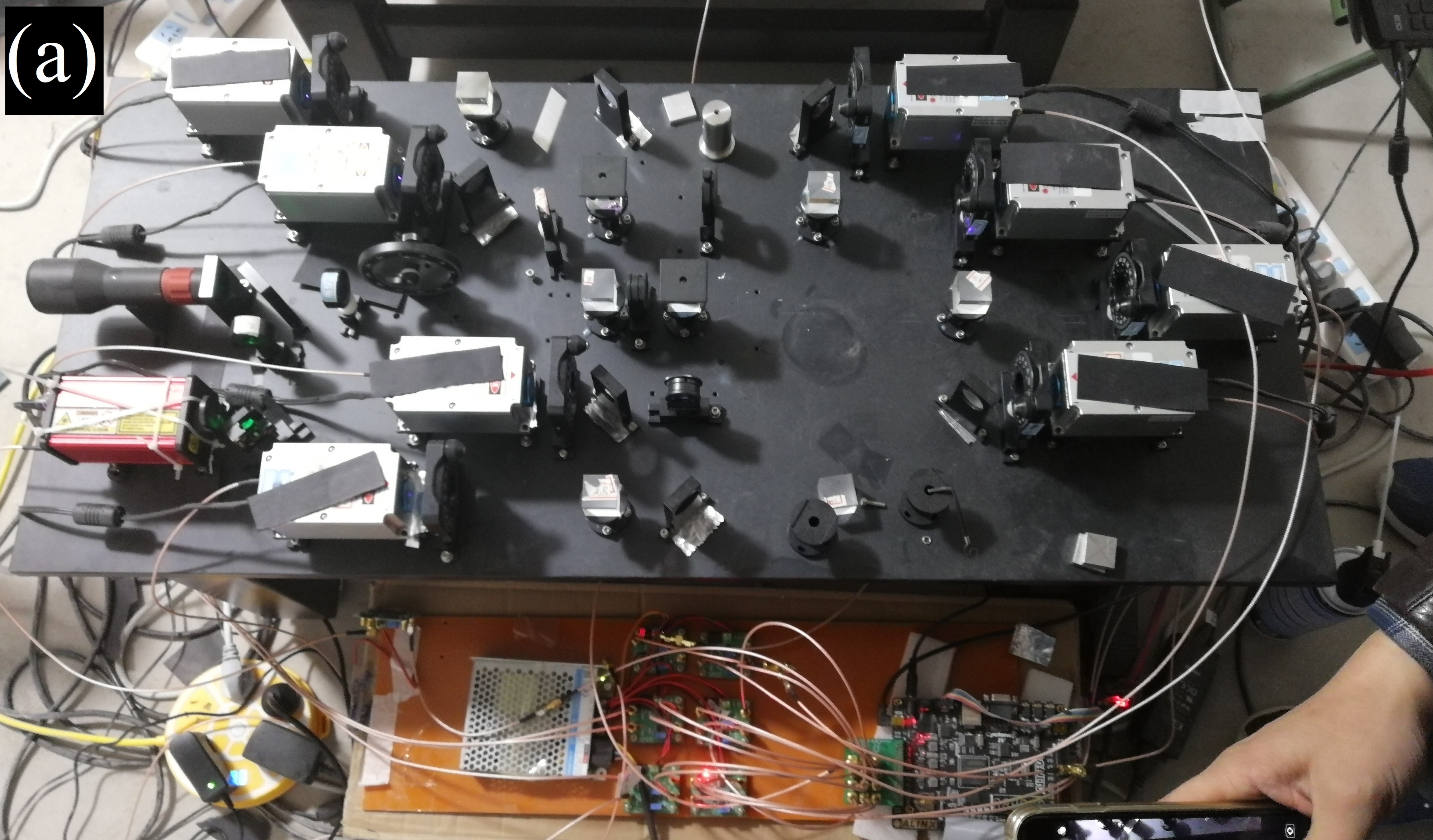}%
		}%
		\mbox{%
			\includegraphics[scale=0.0416]{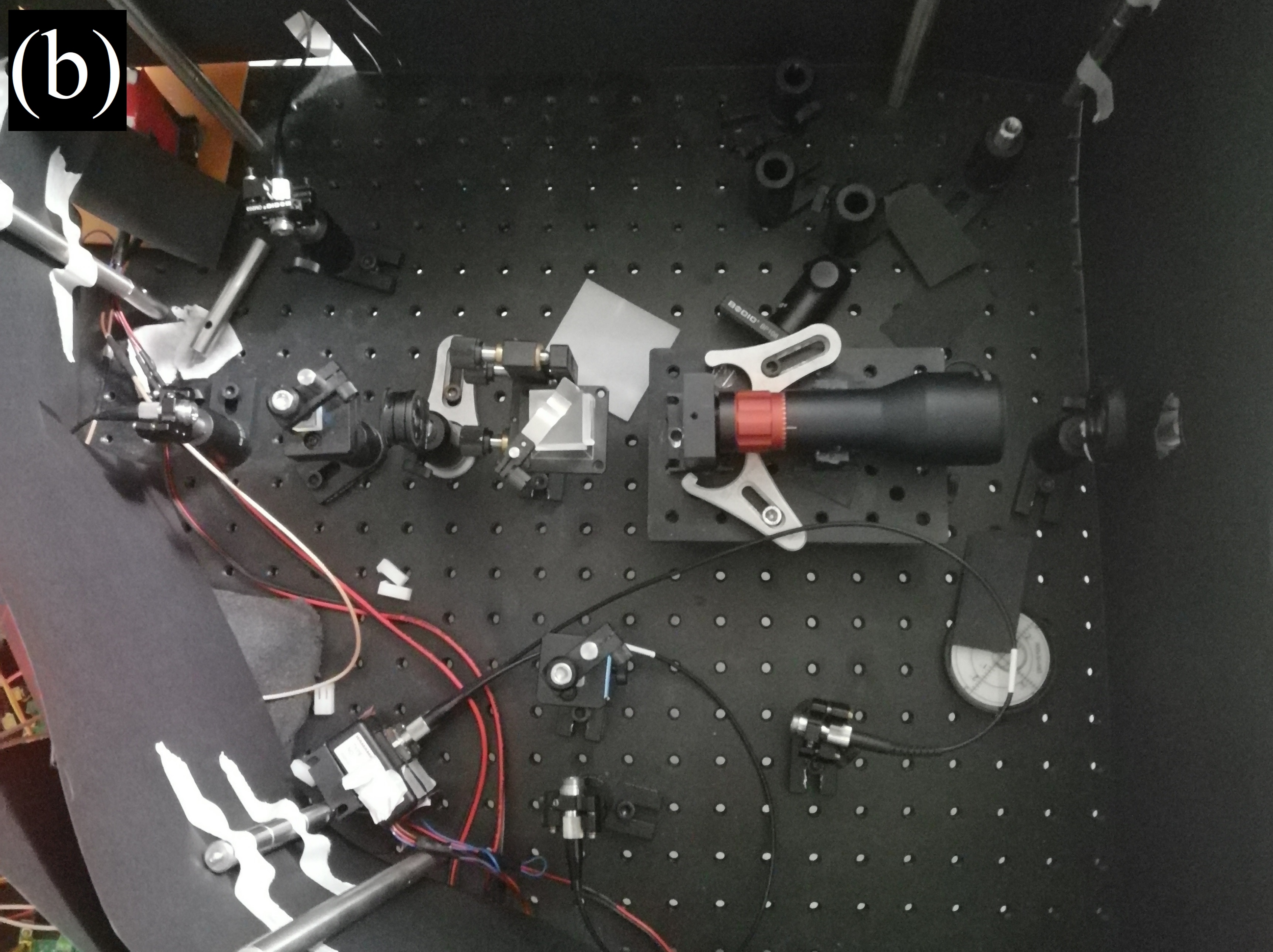}%
		}
		
		\mbox{%
			\includegraphics[scale=0.04]{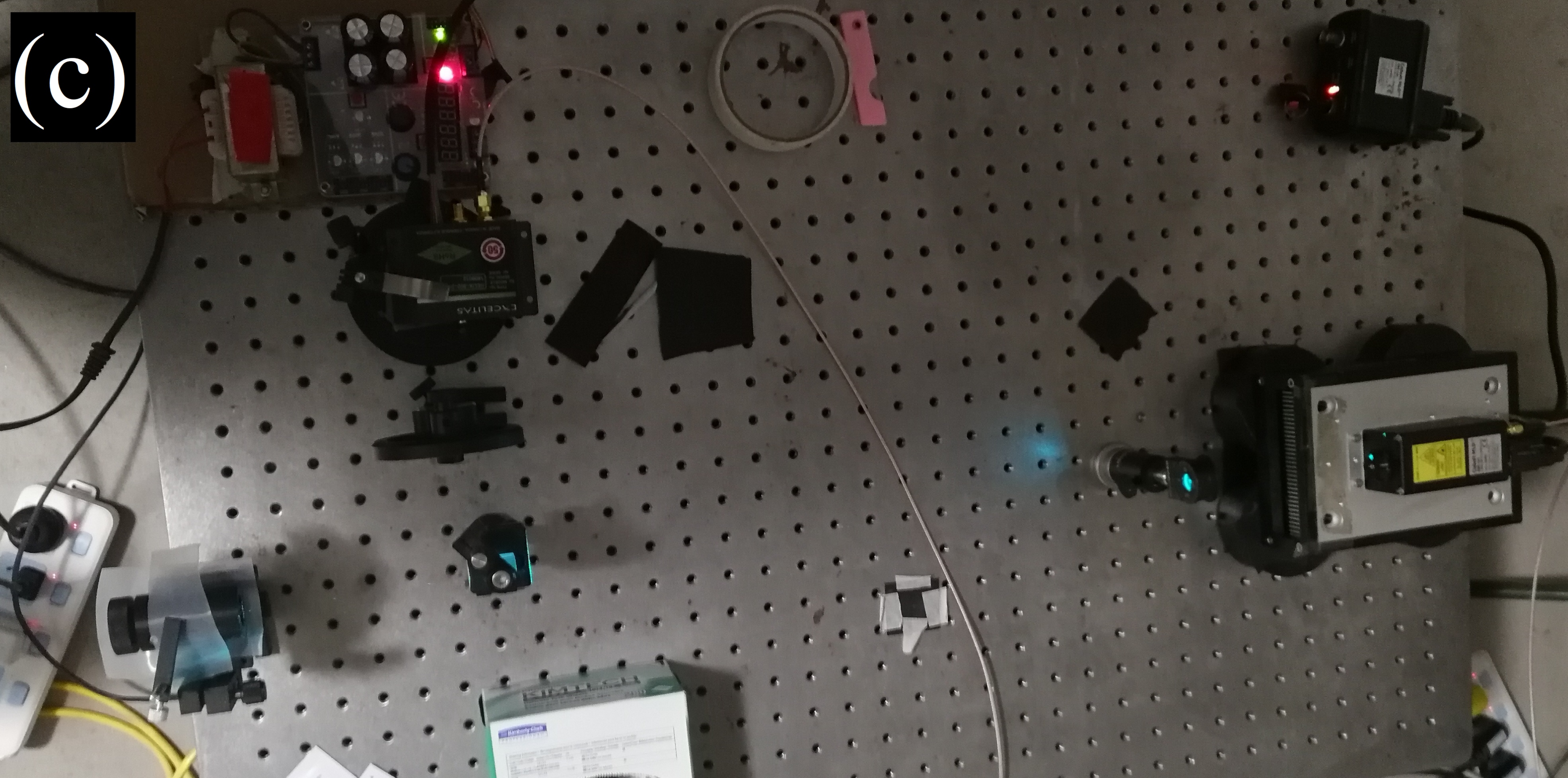}%
		}%
		\mbox{%
			\includegraphics[scale=0.045]{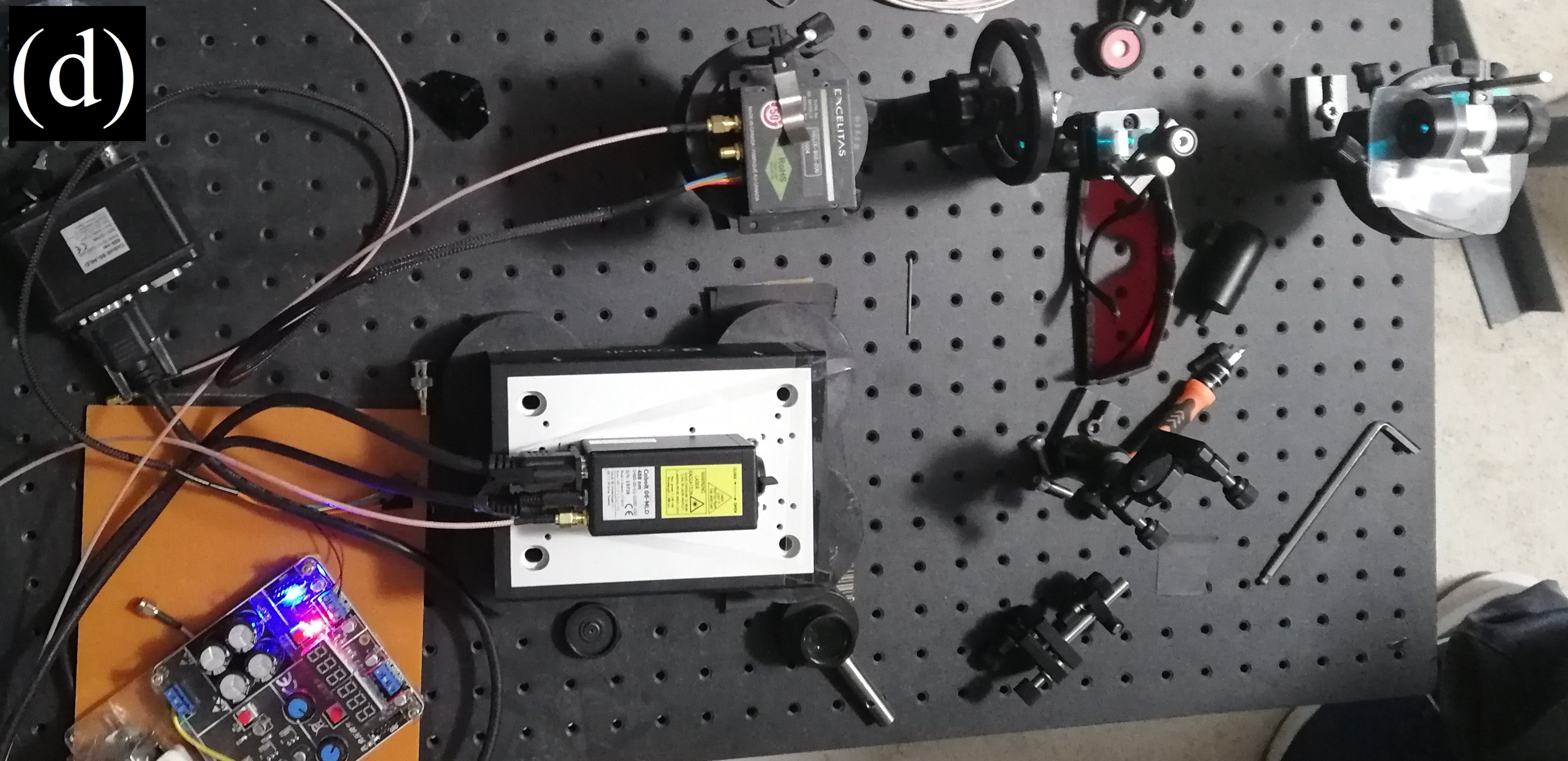}%
		}
		
		\mbox{%
			\includegraphics[scale=0.06]{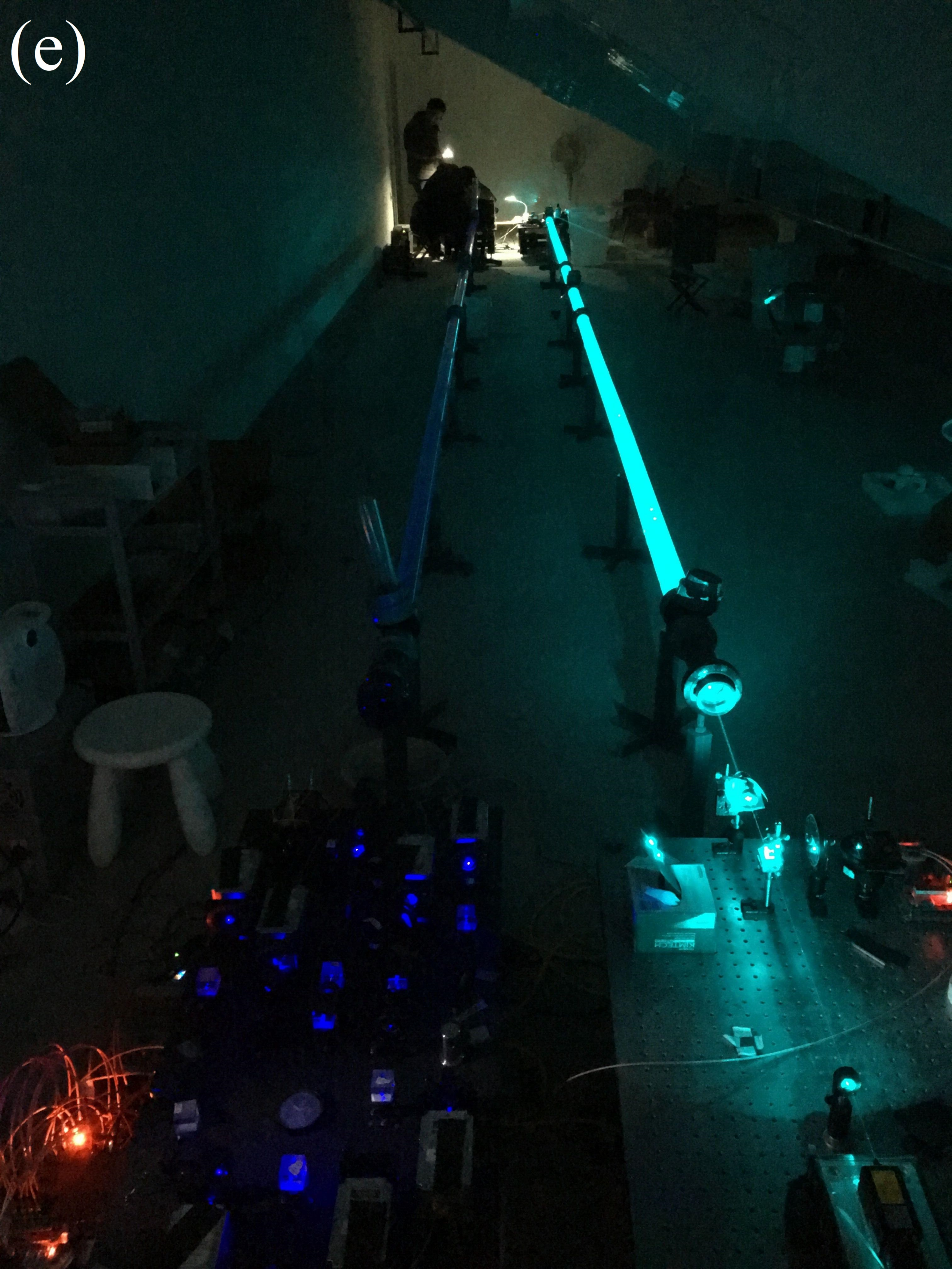}%
			%e)\includegraphics[scale=0.04]{bobclass}%
		}
		
		\caption{\label{fig:epsart3-1} Experimental setups of the UWQKD system. (a)
			The transmitter (Alice). (b) The receiver (Bob). (c) The optical classical communincation setup at the Alice end.
			(d) The optical classical communincation setup at the
			Bob end. (e) Experiment is being performed.}
	\end{figure}


\begin{thebibliography}{24}
		\bibitem[1]{BennettBB84}C. H. Bennett and G. Brassard, "Quantum cryptography: Public key distribution and coin tossing," Theor. Comput.	Sci. 560, 7-11 (2014).
		\bibitem[2]{fiber1}H. J. Kimble, "The quantum internet," Nature 453, 1023-1030 (2008).
		\bibitem[3]{fiber2tokyonetwork}M. Fujiwara, H. Ishizuka, S. Miki, T. Yamashita, Z. Wang, A. Tanaka, K. Yoshino, Y. Nambu, S. Takahashi, A. Tajima,	A. Tomita, and T. Hasegawa, "Field test of quantum key distribution in the tokyo qkd network," Opt. express 19, 10387-10409 (2011).
		\bibitem[4]{fiber3TFqkd}M. Lucamarini, Z. L. Yuan, J. F. Dynes, and A. J. Shields, "Overcoming the rate-distance limit of quantum key distribution without quantum repeaters," Nature 557, 400-403 (2018).
		\bibitem[5]{satalite1}J.-G. Ren, P. Xu, H.-L. Yong, L. Zhang, S.-K. Liao, J. Yin, W.-Y. Liu, W.-Q. Cai, M. Yang, L. Li, H.-Y. Wu, S. Wan,	L. Liu, D.-Q. Liu, Y.-W. Kuang, Z.-P. He, P. Shang, C. Guo, R.-H. Zheng, K. Tian, Z.-C. Zhu, N.-L. Liu, C.-Y. Lu, R. Shu, Y.-A. Chen, C.-Z. Peng, J.-Y. Wang, and J.-W. Pan, "Ground-to-satellite quantum teleportation," Nature 549, 70-73 (2017).
		\bibitem[6]{satalite2}S.-K. Liao, W.-Q. Cai, W.-Y. Liu, L. Zhang, Y. Li, J.-G. Ren, J. Yin, Q. Shen, Y. Cao, Z.-P. Li, F.-Z. Li, X.-W. Chen, L.-H. Sun, J.-J. Jia, J.-C. Wu, X.-J. Jiang, J.-F. Wang, Y.-M. Huang, Q. Wang, Y.-L. Zhou, L. Deng, T. Xi, L. Ma, T. Hu, Q. Zhang, Y.-A. Chen, N.-L. Liu, X.-B. Wang, Z.-C. Zhu, C.-Y. Lu, R. Shu, C.-Z. Peng, J.-Y. Wang, and J.-W. Pan, "Satellite-to-ground quantum key distribution," Nature 549, 43-47 (2017).
		\bibitem[7]{satellite3}H. Takenaka, A. Carrasco-Casado, M. Fujiwara, M. Kitamura, M. Sasaki, and M. Toyoshima, "Satellite-to-ground quantum-limited communication using a 50-kg-class microsatellite," Nat. photonics 11, 502-508 (2017).
		\bibitem[8]{UWQKDtheroeyOE}J. Gariano and I. B. Djordjevic, "Theoretical study of a submarine to submarine quantum key distribution systems," Opt. express 27, 3055-3064 (2019).
		\bibitem[9]{UWQKDtheroeyHINDU}M. Lopes and N. Sarwade, "Optimized decoy state qkd for underwater free space communication," Int. J. Quantum Inf. 16, 1850019 (2018).
		\bibitem[10]{Zhaoshichengbiyelilun}S.-C. Zhao, X.-H. Han, Y. Xiao, Y. Shen, Y.-J. Gu, and W.-D. Li, "Performance of underwater quantum key distribution with polarization encoding," JOSA A 36, 883-892 (2019).
		\bibitem[11]{Jin2020}C.-Q. Hu, Z.-Q. Yan, J. Gao, Z.-M. Li, H. Zhou, J.-P. Dou, and X.-M. Jin, "Decoy-state quantum key distribution over a long-distance high-loss air-water channel," Phys. Rev. Appl. 15, 024060 (2021).
		\bibitem[12]{Jinpolar55m}C.-Q. Hu, Z.-Q. Yan, J. Gao, Z.-Q. Jiao, Z.-M. Li, W.-G. Shen, Y. Chen, R.-J. Ren, L.-F. Qiao, A.-L. Yang, H. Tang, and X.-M. Jin, "Transmission of photonic polarization states through 55-m water: towards air-to-sea quantum	communication," Photonics Res. 7, A40-A44 (2019).
		\bibitem[13]{Jintwist55m}Y. Chen, W.-G. Shen, Z.-M. Li, C.-Q. Hu, Z.-Q. Yan, Z.-Q. Jiao, J. Gao, M.-M. Cao, K. Sun, and X.-M. Jin, "Underwater transmission of high-dimensional twisted photons over 55 meters," PhotoniX 1, 1-11 (2020).
		\bibitem[14]{JinUWQKD3m}L. Ji, J. Gao, A.-L. Yang, Z. Feng, X.-F. Lin, Z.-G. Li, and X.-M. Jin, "Towards quantum communications in free-space seawater," Opt. Express 25, 19795-19806 (2017).
		\bibitem[15]{UWQKDtwistcanada}F. Bouchard, A. Sit, F. Hufnagel, A. Abbas, Y. Zhang, K. Heshami, R. Fickler, C. Marquardt, G. Leuchs, and	E. Karimi, "Quantum cryptography with twisted photons through an outdoor underwater channel," Opt. express 26, 22563-22573 (2018).
		\bibitem[16]{QKDottawariver}F. Hufnagel, A. Sit, F. Grenapin, F. Bouchard, K. Heshami, D. England, Y. Zhang, B. J. Sussman, R. W. Boyd, G. Leuchs, and E. Karimi, "Characterization of an underwater channel for quantum communications in the ottawa river," Opt. express 27, 26346-26354 (2019).
		\bibitem[17]{Zhaoshichengbiyeshiyan}S. Zhao, W. Li, Y. Shen, Y. Yu, X. Han, H. Zeng, M. Cai, T. Qian, S. Wang, Z. Wang, Y. Xiao, and Y. Gu, "Experimental investigation of quantum key distribution over a water channel," Appl. optics 58, 3902-3907 (2019).
		\bibitem[18]{ook1}S. Okada, T. Yendo, T. Yamazato, T. Fujii, M. Tanimoto, and Y. Kimura, "On-vehicle receiver for distant visible light road-to-vehicle communication," in 2009 IEEE intelligent vehicles symposium, (IEEE, 2009), pp. 1033-1038.
		\bibitem[19]{ook2}T. Saito, S. Haruyama, and M. Nakagawa, "A new tracking method using image sensor and photo diode for visible light road-to-vehicle communication," in 2008 10th International Conference on Advanced Communication Technology,	vol. 1 (IEEE, 2008), pp. 673-678.
		\bibitem[20]{Charles1988Privacy}C. H. Bennett, G. Brassard, and J.-M. Robert, "Privacy amplification by public discussion," SIAM journal on Comput. 17, 210-229 (1988).
		\bibitem[21]{Privacy2}C. H. Bennett, G. Brassard, C. Crépeau, and U. M. Maurer, "Generalized privacy amplification," IEEE Transactions on Inf. Theory 41, 1915-1923 (1995).
		\bibitem[22]{zhangchunmei}C.-M. Zhang, X.-T. Song, P. Treeviriyanupab, M. Li, C. Wang, H.-W. Li, Z.-Q. Yin, W. Chen, and Z.-F. Han, "Delayed error verification in quantum key distribution," Chin. science bulletin 59, 2825-2828 (2014).
		\bibitem[23]{toeplitz}H. Krawczyk, "Lfsr-based hashing and authentication," in Annual International Cryptology Conference, (Springer, 1994), pp. 129-139.
		\bibitem[24]{decoyPRA}X. Ma, B. Qi, Y. Zhao, and H.-K. Lo, "Practical decoy state for quantum key distribution," Phys. Rev. A 72, 012326	(2005).
			
		\end{thebibliography}
\end{document}